\documentclass{aa}

\usepackage{graphicx}
\usepackage{orcidlink}
\usepackage{txfonts}
\usepackage{xcolor}
\usepackage[]{hyperref}
\hypersetup{
    colorlinks=true,
    urlcolor=blue,
    linkcolor=red,
    breaklinks=true,
    citecolor=blue
}

\usepackage{float}
\usepackage{afterpage}
\usepackage{threeparttable}

\def\HI{H{\sc i}\,}

\def\Ms{$\textrm{M}_{\odot}$}

\begin{document}

   \title{Giant Low-Surface Brightness Galaxies in the Global Picture of Galaxy Formation and Evolution}
   \subtitle{Scaling Relations of Giant Galactic Disks from a Statistically-Significant Sample}

   \author{Fedor M. Kolganov\inst{1,2}\orcidlink{0000-0002-9609-7980}
          \and
          Igor V.  Chilingarian\inst{4,3}\orcidlink{0000-0002-7924-3253}
           \and
          Anna S. Saburova\inst{3}\orcidlink{0000-0002-4342-9312}
          \and
           Damir Gasymov\inst{5,3}\orcidlink{0000-0002-1750-2096}
           \and
           Evgenii V. Rubtsov\inst{3}\orcidlink{0000-0001-8427-0240}
          \and
          Vladislav I. Klochkov\inst{6}\orcidlink{0000-0003-3095-8933}
           \and
          Kirill A. Grishin\inst{7,3}\orcidlink{0000-0003-3255-7340}
          \and
          Mariia  Demianenko\inst{8, 9}\orcidlink{0000-0002-8297-6386}
          }

   \institute{
            Felix Bloch Institute for Solid State Physics, Leipzig University, Linnéstraße 5, 04103 Leipzig, Germany\\
              \email{kolganov@voxastro.org} \and
            Faculty of Physics and Earth System Sciences, Leipzig University, Linnéstraße 5, 04103 Leipzig, Germany
         \and
             Sternberg Astronomical Institute, Moscow M.V. Lomonosov State University, Universitetskij pr., 13,  Moscow, 119234, Russia
        \and
           Center for Astrophysics --- Harvard and Smithsonian, 60 Garden Street MS09, Cambridge, MA 02138, USA
        \and
            Astronomisches Rechen-Institut, Zentrum f\"ur Astronomie der Universit\"at Heidelberg, M\"onchhofstr.\ 12--14, 69120 Heidelberg, Germany
        \and
        Institute of Astronomy, Russian Academy of Sciences, 119017, Moscow, Russia
        \and
            Observatoire de Paris, LUX, CNRS, PSL University, F-75014 Paris, France
        \and
            Max-Planck-Institut für Astronomie, Königstuhl 17, 69117 Heidelberg, Germany
        \and   
            Department for Physics and Astronomy, Heidelberg University, Im Neuenheimer Feld 226, 69120 Heidelberg, Germany 
             }

   \date{Received 2026; accepted}

 
  \abstract
{
Giant low-surface brightness galaxies (gLSBGs) challenge standard merger-driven galaxy assembly models due to their large, dynamically cold disks, with radii often larger than 100~kpc and dynamical masses exceeding ${\sim}10^{12}$~\Ms. We identify $60$ gLSBGs, including a volume-complete subsample of $35$ galaxies out to $z=0.1$ in the 120~sq.deg. area. Using new and archival photometric and spectroscopic data, we homogeneously analyze their structural properties and derive central velocity dispersions. We update the gLSBG volume density estimate in the local Universe to $(3.6 \pm 0.6) \cdot 10^{-5}~\text{Mpc}^{-3}$. This number corresponds to about 1 in 4000 galaxies in the $g$-band luminosity range $0.14-1.37 \cdot 10^{11}~L_{\odot}$ out to $z=0.1$, which is consistent with EAGLE cosmological simulations but 3.8 times lower than the corresponding value in TNG100. Central surface brightness and scalelength of the gLSB disks scale with the luminosity of the central component, becoming fainter and flatter as the central component grows more luminous, suggesting rapid disk formation rather than the inside-out growth typical of high surface brightness late-type galaxies. Many gLSBGs also show extended UV counterparts in GALEX, strengthening their connection to XUV-disk galaxies. Finally, comparison with superluminous spirals shows that even complete gas-to-star conversion cannot transform a typical gLSBG into a superspiral, owing to their lower total baryonic mass.

  }


   \maketitle
\nolinenumbers

%
\section{Introduction}
A class of low-surface brightness galaxies (LSBGs), stellar systems with central surface brightness values fainter than $22~\text{mag arcsec}^{-2}$ in the $B$ band, includes galaxies of various morphological types ranging from irregular dwarfs to the most extended disks  in the Universe~\citep{Boissier2016}. Giant low-surface brightness galaxies (gLSBGs) reside in the high-mass end of this class: the radii of their disks  in the optical spectral domain are larger than 50~kpc.\footnote{We define the radius as either $4h$, where $h$ is the exponential scalelengths in \textit{g}-band or radius of the $28~\text{mag arcsec}^{-2}$ isophote in the $B$ band, whichever is larger.} They often show extended \HI counterparts of similar size \citep{Pickering1997, Lelli2010, Mishra2017}. 

Large gaseous disks  with little stellar content make gLSBGs the most direct probes of the dark matter distribution out to large radii. That is why, since the discovery of the prototype Malin~1 \citep{Bothun1987}, significant effort has been dedicated to identifying analogous systems. However, until recently, the search for gLSBGs was strongly limited by the depth of available imaging surveys. This resulted in a small number of known systems, as their faint disks  were often lost in the noise of the sky background, which also made gLSBGs appear as early-type galaxies \citep[][]{Hagen2016} and often classified as such in large galaxy catalogs.

The question of how gLSBGs were formed still remains a matter of debate \citep[see, e.g.][]{Lelli2010, Galaz2015, Boissier2016, Hagen2016, saburovaetal2021, Junais2024, saburovaetal2026}. In particular, \citet{saburovaetal2021} suggested that multiple formation channels may exist resulting in an inhomogeneous galaxy class; however, their conclusions were based on a small sample of seven gLSBGs.


The advent of deep, wide-field photometric surveys now enables a systematic, sample-based approach to studying gLSBGs, which we adopt in this work.
Our sample of gLSBGs comprises (i) a core subsample which is complete within a 100~Mpc$^3$ co-moving volume discussed in \citep{vol_dens2023} and (ii) additional gLSBGs serendipitously discovered by our team across other sky regions. For all galaxies in the sample here we perform homogeneous analysis of their surface brightness profiles and derive structural parameters for their disk and bulge components. By placing these parameters, alongside stellar velocity dispersion estimates, onto standard scaling relations and comparing them with other galaxy populations, we draw conclusions regarding the formation histories and overall homogeneity and properties of gLSBGs as a class.


Understanding which formation channels produce such systems in the $\Lambda$CDM paradigm requires determining how frequently they appear in the Universe. \citet{vol_dens2023} laid the groundwork for this by analyzing a 120~deg$^2$ field in Subaru Hyper Suprime-Cam (HSC) data, showing that gLSBGs are less rare than previously assumed and broadly consistent with predictions from the EAGLE cosmological hydrodynamical simulation \citep{2015MNRAS.446..521S}. Here we capitalize on this approach by completing the sample of gLSBG candidates within the same field by (i) accurately re-measuring their sizes and (ii) obtaining the spectroscopic redshifts for the candidates without such measurements, thus creating a complete volume limited sample of gLSBGs.

We use AB magnitudes throughout the paper \citep{1983ApJ...266..713O}. We assumed the following cosmological parameters: $H_0=67.4$~km~s$^{-1}$~Mpc$^{-1}$ and $\Omega_m=0.315$, as taken from \citep{2020A&A...641A...6P}.

The paper is organized as follows. In Sect.~\ref{data_obs}, we describe the sample of gLSBG candidates and the long-slit spectral observations and data reduction. The analysis of spectral and photometric data is described in Sect.~\ref{data_analysis}. We present the main results in Sect.~\ref{results} and discuss them in Sect.~\ref{discussion}. We give the conclusions in Sect.~\ref{conclusions}.

\begin{figure}
    \centering
    \includegraphics[width=\hsize]{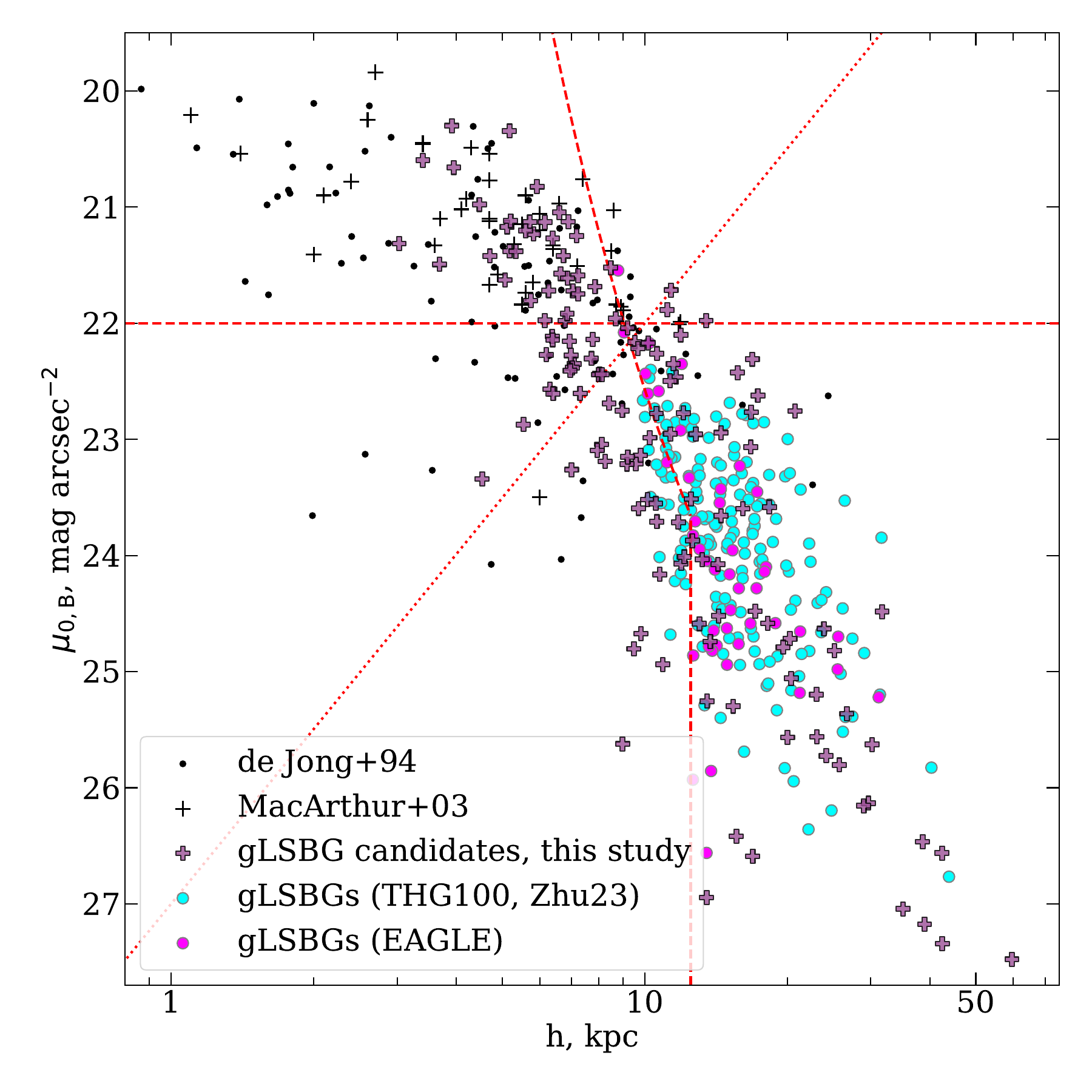}
    \caption{The relation between the B-band disk central surface brightness and the exponential scale length as was originally introduced by \citet{Sprayberry1995}. Squares represent literature data, while crosses indicate all gLSBGs candidates identified in this work, including those not included in the final sample. Yellow and blue symbols show galaxies from the TNG100 \citep{Zhu23} and EAGLE \citep{vol_dens2023} simulations, respectively. The diagonal line marks the diffuseness index criterion; galaxies located below this line satisfy the selection. The selection adopted in this work occupies the bottom-right rectangle of the diagram.}
    \label{fig:data_sample}
\end{figure}
\section{Galaxy sample, observations, archival data}\label{data_obs}

Our core subsample consists of gLSBG candidates in the equatorial HSC field \citep{vol_dens2023}, independently identified through visual inspection by our team. To achieve volume completeness, this sample combines objects at previously known redshifts with newly identified candidates in the same footprint lacking redshift data. We complement this core set with additional candidates serendipitously discovered across other sky regions, with and without redshift information from the literature.

\begin{figure}
    \centering
    \includegraphics[width=\hsize]{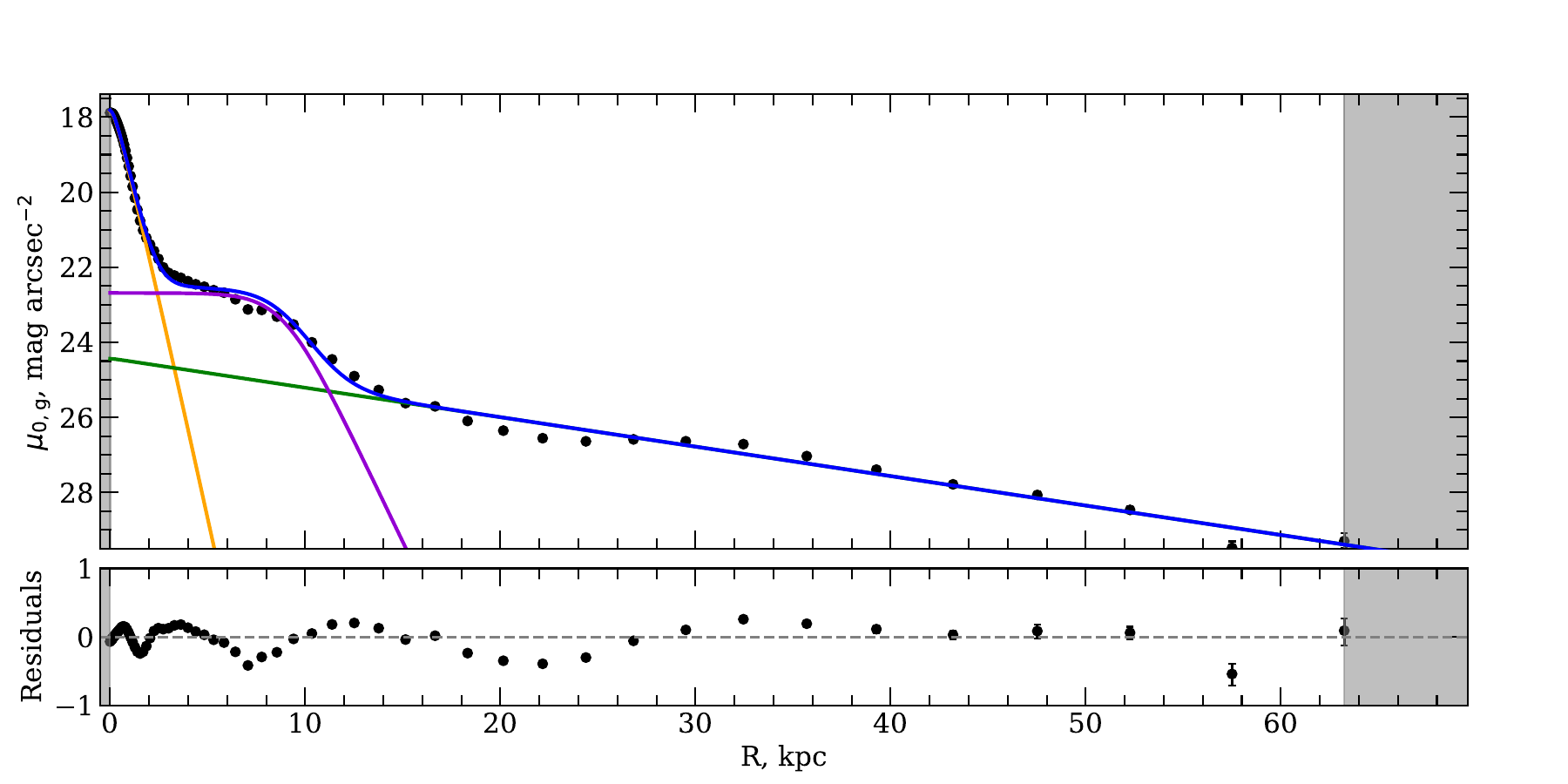}
    \includegraphics[width=\hsize]{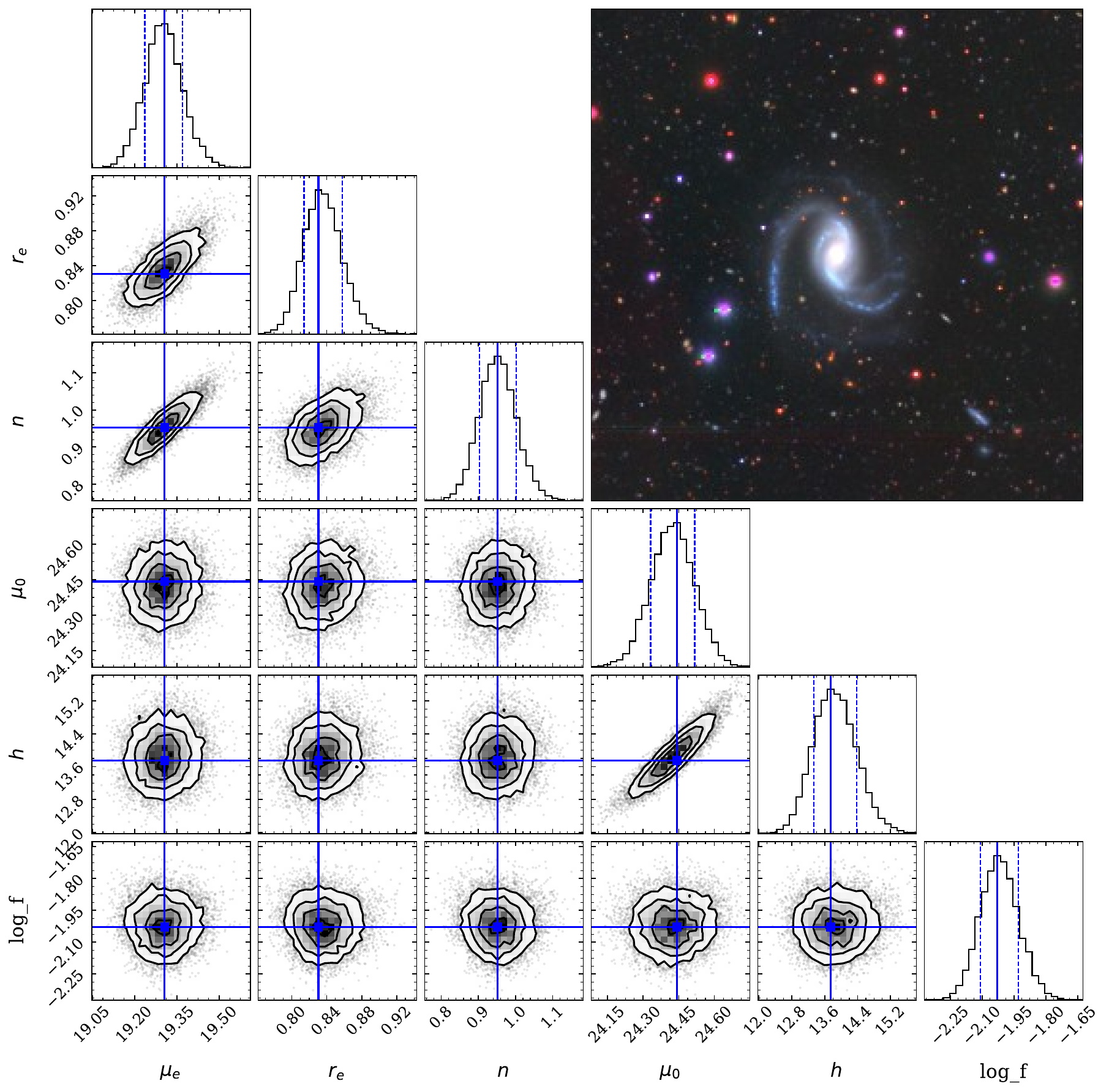}
    \caption{Example of the fit to the {\it g}-band surface brightness profile of Mrk~592 (top panel). Black circles represent the observed profile, while the blue line shows the total model. The green and yellow lines indicate the contributions from the disk and the S\'ersic bulge components, respectively, and the violet line corresponds to the ring component. The lower sub-panel displays the residuals of the fit. The bottom panels present the MCMC corner plots for the fitted parameters.}
    \label{fig:fit}
\end{figure}

We performed surface photometry and light profile decompositions for all 148 gLSBG candidates to determine their classifications and homogeneously derive their disk and bulge structural parameters (see Fig. \ref{fig:data_sample} for an overview of the results and Fig.~\ref{fig:fit} for a decomposition example).

We carried out spectroscopic observations to determine the redshifts of the candidates where those values were not available in the literature. We used the ESO Faint Object Spectrograph and Camera v.2 \citep[EFOSC2;][]{efosc2_paper}; FAst Spectrograph for the Tillinghast Telescope \citep[FAST;][]{fast_paper} and Transient Double-beam Spectrograph~\citep[TDS;][]{tds_paper} (see the observational log in Table \ref{tab:obslog}). We also used new and archival spectral data to estimate stellar velocity dispersion for a subset of confirmed gLSBGs to compare them with galaxies of other types. Below, we provide the details of our new spectroscopic observations. 

\begin{table}
\caption{Log of spectroscopic observations.}
\label{tab:obslog}
\centering
\begin{threeparttable}
\addtolength{\tabcolsep}{-0.2em}
\begin{tabular}{l c c c}
\hline\hline
Object & RA & Dec & $t_{\rm exp}$, s \\
\hline
\multicolumn{4}{c}{EFOSC2 / FAST} \\
\hline
NGC7478 & 346.2359 & 2.5778 & 2400 / 1800 \\
J020825.73-031427.6 & 32.1071 & -3.2411 & 2400 / 1500 \\
J020855.89+024610.9 & 32.2329 & 2.7697 & 2400 / 1800 \\
J021122.24-032714.3 & 32.8427 & -3.4540 & 2100 / 1200 \\
J021310.77-023444.5 & 33.2949 & -2.5790 & 2700 / 1800 \\
J023923.45-034353.9 & 39.8477 & -3.7317 & 2400 / 1800 \\
J223149.54-005321.5 & 337.9564 & -0.8893 & 2400 / 1800 \\
J223757.77+020943.8 & 339.4907 & 2.1622 & 1800 / 1200 \\
J224033.18+013725.8 & 340.1382 & 1.6238 & 2400 / 1200 \\
J224841.04+022413.6 & 342.1710 & 2.4038 & 2400 / 1200 \\
J235920.45+015556.6 & 359.8352 & 1.9324 & 2400 / 1800 \\
\hline
\multicolumn{4}{c}{EFOSC2} \\
\hline
J004625.86-142345.5$^\dagger$ & 11.6078 & -14.3959 & 3600 \\
J022551.78-031813.0 & 36.4658 & -3.3036 & 2400 \\
J223730.20+022719.7 & 339.3758 & 2.4555 & 2400 \\
J223845.81+022119.2 & 339.6908 & 2.3554 & 4000 \\
J231938.79+022300.3 & 349.9116 & 2.3838 & 2400 \\
J234407.50+010559.5 & 356.0313 & 1.0999 & 4500 \\
LEDA1028220 & 29.8268 & -6.8240 & 2400 \\
LEDA1053433 & 28.5511 & -4.7906 & 2400 \\
LEDA1071586 & 36.8263 & -3.4159 & 2400 \\
\hline
\multicolumn{4}{c}{FAST} \\
\hline
J022033.54-062602.5 & 35.1398 & -6.4340 & 1800 \\
J022337.32-025514.9 & 35.9055 & -2.9208 & 1800 \\
J023306.58-032142.8 & 38.2774 & -3.3619 & 1800 \\
J221845.77+013215.4 & 334.6907 & 1.5376 & 1800 \\
J223718.24+012037.3 & 339.3260 & 1.3437 & 1200 \\
J223728.22+022722.3 & 339.3676 & 2.4562 & 1800 \\
J225258.42+023742.2 & 343.2434 & 2.6284 & 1800 \\
LEDA768523$^\dagger$ & 67.5223 & -26.3743 & 1800 \\
\hline
\multicolumn{4}{c}{TDS} \\
\hline
J002814.67+300618.2$^\dagger$ & 7.0611 & 30.1051 & 3600 \\
J173451.16+250451.8$^\dagger$ & 263.7132 & 25.0811 & 3600 \\
UGC11596$^\dagger$ & 308.9355 & 1.3675 & 3600 \\
\hline
\end{tabular}
\begin{tablenotes}
  \small
  \item[$\dagger$] Photometry from Legacy Survey dr9; the remaining objects are from HSC DR2 Wide.
\end{tablenotes}
\end{threeparttable}
\end{table}

\subsection{Long-slit spectroscopic observations and data reduction}
We conducted spectroscopic observations of 31 candidate gLSBGs without previously known redshifts at three ground-based observatories in long-slit regime. For the observation log see Table \ref{tab:obslog}.

\subsubsection{EFOSC2} \label{sect:efosc2}

Observations of 22 galaxies were conducted with the ESO Faint Object Spectrograph and Camera v.2 \citep{efosc2_paper} at the 3.5-m ESO New Technology Telescope during the observing run in October 2022 as part of the program "Redshift confirmation of giant low-surface brightness galaxies and their compact elliptical satellites" (Program ID: 110.246F; P.I. KG). We opted for a low-resolution setup with the grism Gr\#4 and 2$\times$2 binning to increase the signal-to-noise ratio per pixel. We used a 1~arcsec-wide 4.1~arcmin-long slit oriented along the major axes of the galaxies. The spectra cover the wavelengths from 4085 to 7520~\AA\ and have resolving power $\text{R} \simeq 1100$. During each night, at least one of the two standard stars, Hiltner~600 or Feige~110, was observed. 

We determined the spectral line spread function (LSF) as a function of wavelength by fitting Gaussian profiles to emission lines of sufficient intensity in the arc lamp spectra, as no significant higher-order Gauss–Hermite ($h_3, h_4$) deviations were detected.

The spectral line-spread-function (LSF) rapidly decreases along the wavelength from its maximum value of $\sigma_{\rm inst}$=370~km/s at 4100~\AA, reaching a minimum of 200~km/s at 6500~\AA, and rises to 225~km/s at 7500~\AA. We approximated it using a second degree polynomial as shown below:
\begin{equation}
    \sigma(\lambda) = 1419 - 3.75\cdot10^{-1}\lambda + 2.89\cdot10^{-5}\lambda^2,
\end{equation}
where $\sigma$ is in km/s and $\lambda$ is in \r{A}.

The absolute values of the single-line corrections to the zeropoint of the wavelength solution (see Sect. \ref{spectral_red}) measured from night sky airglow lines did not exceed 4~\AA, which is smaller than the spectral LSF width at its minimum. 

\subsubsection{FAST} \label{sect:fast}

19 gLSBG candidates were observed with the FAst Spectrograph for the Tillinghast Telescope \citep{fast_paper} at the Fred Lawrence Whipple Observatory in Arizona, USA, in the 2022B period; program 235 "Giant Low-Surface Brightness Galaxies" P.I. IC). Similar to the EFOSC2 setup, we opted for the lowest resolution grating with 300 lines $\text{mm}^{-1}$ and a 3 arcsec wide and 3 arcmin long slit. This setup provided mean resolving power $\text{R} \simeq 2500$ and spectral coverage from 3475 to 7415~\AA. A number of standard stars were observed. For the purposes of flux calibration and telluric correction, we used spectra of the following ones: Feige110, BDp284211, BDp284211, G191B2B, Feige34, and Hiltner600.

The bias-subtracted, flat-fielded and wavelength-calibrated spectra were provided by the Special Astrophysical Observatory Telescope Data Center. We used our set of software tools described in Sect.~\ref{spectral_red} to finalize the reduction, by applying flux calibration, telluric correction, subtracting the sky background, cleaning each frame from cosmic ray hits, combining individual exposures, and performing optimal extraction of 1D spectra.

For the calculation of the instrumental function of the FAST setup, we used the foreground spectrum of the night sky. The measured resolution of spectra smoothly decreases from 160 km/s at 3500~\AA\ to 75~km/s at 7400~\AA. We approximated it using a second degree polynomial as shown below:
\begin{equation}
    \sigma(\lambda)= 151 + 1.32\cdot10^{-2}\lambda - 3.24\cdot10^{-6}\lambda^2,
\end{equation}
where $\sigma$ is in km/s and $\lambda$ in \r{A}.

\subsubsection{TDS} \label{sect:tds}

Three additional galaxies were observed with the Transient Double-beam Spectrograph~\citep{tds_paper} mounted on the 2.5-meter telescope at the Caucasus Mountain Observatory, Russia in July 2023. The instrument features two wavelength channels separated by a dichroic beam splitter with 50\% transmission at 5740~\AA. The blue channel covers 3600--5770~\AA\ with a resolving power of $R \simeq 1300$, while the red channel covers 5670--7460~\AA\ with $R \simeq 2500$. The red channel used the permanently installed R grating with a ruling density of 1200~lines~$\text{mm}^{-1}$, whereas for the blue channel, we selected the lower-resolution B grating with 900~lines~$\text{mm}^{-1}$. We used a 1~arcsec-wide, 3~arcmin-long slit.

We obtained the line spread function of the setup using ARC lamp spectra by applying the procedure described in \ref{sect:efosc2}. In the blue channel, the dispersion steadily decreases from 130~km/s at 3600~\AA\ to 85~km/s at 5700~\AA, dropping by 25~km/s at the beginning of the span of the red channel, and levels out at 40~km/s at 6800~\AA. We approximated it using a piece-wise second degree polynomial as shown below:
\begin{equation}
    \sigma^{\text{B}}(\lambda) = 375 - 9.78\cdot10^{-2}\lambda - 8.35\cdot10^{-6}\lambda^2
\end{equation}
\begin{equation}
    \sigma^{\text{R}}(\lambda) = 608 - 1.61\cdot10^{-1}\lambda - 1.14\cdot10^{-5}\lambda^2,
\end{equation}
where $\sigma$ is in km/s and $\lambda$ is in \r{A} with B and R superscripts denoting blue and red channels respectively.

We reduced data from each channel independently, as the dichroic splitter feeds physically separate optical paths with dedicated CCD detectors. We then stitched the extracted 1D spectra onto a common wavelength grid by oversampling the blue channel spectrum to match the red channel dispersion, preserving all spectral information. In the overlap region, we masked the blue channel spectrum, retaining only the red channel data.

\subsection{Spectroscopic data reduction}\label{spectral_red}

We reduced our follow-up spectroscopic observations with the {\sc idl} data reduction pipeline originally designed for the Transient Double-beam Spectrograph, but applicable to any long-slit spectral data. We previously applied it to the IDS \citep{2025A&A...702A..42G} and DEIMOS \citep{Saburova2024ApJ...973..167S} long-slit spectra. The procedure is briefly described in \citet{2018ApJ...858...63C}. Due to the uniformity of the spectroscopic Line Spread Function along the slit and the slit itself being substantially larger than the object sizes in all three cases, we approximated the sky background by averaging the outer parts of the slit and then subtracted this averaged sky vector from every position on the slit. In addition to the wavelength calibration based on the ARC spectra we applied (i) a single shift based on a bright atmospheric spectral line to address systematic uncertainty caused by temporal variations in wavelength solution between the science and arc lamp exposures, as well as (ii) barycentric correction. We corrected the spectra for the atmospheric telluric absorption using a technique described in detail in \citet{2023ApJS..266...11B}. We then extracted 1D spectra using the technique described in \citep{Horne_1986}.

\subsection{Archival spectroscopic data}
For the spectroscopic analysis of the subsample with previously known redshifts 
we used archival spectral data from SDSS DR16 \citep[][53 objects]{2020ApJS..249....3A}, DESI DR1 \citep[][42 objects]{Dey2019}, and 6dFGS \citep[][12 objects]{2004MNRAS.355..747J} spectroscopic surveys.

\section{Data analysis}\label{data_analysis}
\subsection{Spectroscopic data analysis using full spectrum fitting}

The 1D spectra of a sample of 31 candidate gLSB galaxies (see Sect.~\ref{data_obs}), obtained from both new observations and archival data, were analysed using the {\sc NBursts} software package \citep{Chilingarian2007a, Chilingarian2007b}.
This package implements a full-spectrum fitting technique to recover the line-of-sight velocity distribution (LOSVD; i.e., radial velocity and velocity dispersion), taking into account the line spread function (LSF) of the spectrograph, as well as the properties of the stellar component (age and metallicity) and the ionized gas (emission-line fluxes).
To model the stellar component, we used the PEGASE.HR~\citep{LeBorgneetal2004} simple stellar population (SSP) models computed for the Salpeter initial mass function \citep[IMF,][]{Salpeter1955}, covering the 3900--6800~\AA\ in the rest frame at a spectral resolution of $R = 10000$ for SDSS, DESI, and 6dFGS archival spectra. For the observational data from EFOSC2, TDS, and FAST, we used MILES SSP models \citep{2015MNRAS.449.1177V} with wavelength range 3500--7400~\AA\ and FWHM~$= 2.51$~\AA.
To model emission lines in this wavelength range we assumed shared kinematic parameters of Gaussian profiles within a single template. In individual cases we used additional emission line templates to describe ambiguities like double peak lines or broad line components.

\subsection{HSC and Legacy Surveys: photometric data}
To analyze the archival photometric data, we first constructed isophotal models for each candidate following the procedure in \citet{vol_dens2023}, correcting for galaxy inclination, Galactic extinction, and cosmological surface brightness dimming. Point spread function (PSF) models for each object were generated using {\sc PSFeX} \citep{2011ASPC..442..435B}. We then performed 1D photometric decompositions on the $g$-band light profiles of all candidates, as well as on the $r$-band profiles for targets with new spectroscopic observations (see Table~\ref{tab:obslog}). Profiles were fitted with (i) a S\'{e}rsic bulge and an exponential disk, alongside (ii) secondary disk, Gaussian and/or Fermi-Dirac components to account for light profile ambiguities, if necessary.

Prior to fitting, we preprocessed each photometric profile by excluding the innermost isophotes within the PSF half-width at half-maximum (HWHM). To avoid contamination from sky-noise-dominated regions, we truncated the outer profile at the turnover radius, defined as the point where the derivative of a second-order polynomial fitted to the outermost 10\% of the profile changes sign.

\subsubsection{Initial bulge+disk decomposition}
Initial parameter guesses for the S\'{e}rsic bulge and exponential disk components are set as follows. The S\'{e}rsic index is initially fixed at $n = 3$, while the effective radius ($r_{1/2}$) and central surface brightness are initialized using the curve-of-growth half-light radius and the surface brightness at that radius. The exponential disk parameters are initialized via a linear fit to the logarithmic light profile between $3\,r_{1/2}$ and $10\,r_{1/2}$ (falling back to the outermost radius with isophotes brighter than 27~mag~arcsec$^{-2}$ if fewer are detected beyond $3\,r_{1/2}$). 

Using least-squares minimization, we then fit a PSF-convolved bulge+disk model with all S\'{e}rsic and disk parameters set free. To ensure robustness, the pipeline employs automated fallback mechanisms: if the minimizer fails to converge or yields a large central brightness drop, the initial $\mu_0$ guesses are iteratively adjusted and parameter boundaries are restricted until a stable baseline fit is achieved.

\subsubsection{Detection and fitting of bars and rings}
We analyze the residuals between the preliminary model and the observed profile at radii $r > 10\,r_{1/2}$ (or $r > 3\,r_{1/2}$ if $10\,r_{1/2}$ lies beyond the $26.5~\text{mag~arcsec}^{-2}$ isophote), restricting the analysis to regions brighter than $26.5~\text{mag~arcsec}^{-2}$. Local residual maxima are identified using the \textsc{find\_peaks} function from \textsc{SciPy} \citep{scipy}, requiring a minimum peak height of $0.15~\text{mag~arcsec}^{-2}$ and a minimum separation of 3~isophotes. 

Each detected peak is assigned an additional model component. If the innermost peak lies within $\max(5\,r_{1/2},\,2h)$, where $h$ is the preliminary disk scalelength, it is modeled with a Fermi--Dirac step-edge profile \citep[][see Appendix~\ref{app:formulas}]{2001A&A...367..405P} and interpreted as a bar feature. Initial parameters are set to the peak radius, the observed surface brightness at that radius, and a decay width of $r_\mathrm{bar}/100~\text{kpc}$ to enforce a sharp boundary. We adopt a Fermi--Dirac function rather than a high-$n$ S\'{e}rsic profile to avoid parameter degeneracies with the central bulge, as only the primary bulge and disk components are evaluated in subsequent analyses. 

All remaining peaks are modeled using Gaussian profiles to capture rings or prominent spiral arm features \citep{2016ApJS..222...10S, 2016PASA...33...62C}, initialized with a width of $1~\text{kpc}$ and peak position and brightness derived directly from the profile at that radius. We then refit the complete multi-component model (bulge + disk + bar/rings) while holding the exponential disk scalelength fixed and allowing all other parameters to vary. To prevent overfitting, any Gaussian component with a fitted FWHM spanning fewer than 3~isophotes is flagged as unresolved and removed. The outcome of this step is saved as an intermediate solution. 

We emphasize that these auxiliary components are not strictly interpreted as definitive physical bars or rings; rather, they are introduced to capture complex features in extended galaxy profiles, mitigating underfitting and ensuring robust parameter estimation for the underlying S\'{e}rsic and exponential disk components.

\subsubsection{Detection and fitting of an outer disk}
We inspect the residual profile following subtraction of the intermediate model (including bar and ring components) at large radii to determine whether an outer disk component is required. A second exponential disk is added if at least three data points in the outer profile lie more than $0.2~\text{mag~arcsec}^{-2}$ above the model fit and isophotes fainter than $26~\text{mag~arcsec}^{-2}$ are detected. Initial parameters for the outer disk are estimated via a linear fit to the surface brightness profile beyond the $26~\text{mag~arcsec}^{-2}$ isophote.

We reject the double-disk model and restore the intermediate single-disk solution if any of the following criteria are met: (i) the inner disk central surface brightness is fainter than $27~\text{mag~arcsec}^{-2}$ ($\mu_{0,\mathrm{inner}} > 27~\text{mag~arcsec}^{-2}$); (ii) the central surface brightness of the inner disk is fainter than that of the outer disk ($\mu_{0,\mathrm{inner}} > \mu_{0,\mathrm{outer}}$); or (iii) the scalelengths of the two disks differ by less than $3~\text{kpc}$. The conditions (i) and (ii) remove models where the inner disk becomes undetectable and a single disk model is preferred. The condition (iii) removes models where both disks fit effectively the same component.

\subsubsection{Final refinement of initial parameter estimates}
We conclude the initialization pipeline with a final least-squares pass to adjust component surface brightnesses and refine the baseline fit. During this optimization step, the S\'{e}rsic index is held fixed at its current best-fit value. 

If a two-disk solution is accepted, only the outer disk central surface brightness is allowed to vary; its scalelength, all bulge and inner disk parameters, and the structural positions and widths of any bar or ring components remain fixed, with only their surface brightnesses left free. Conversely, if a single-disk solution is retained, the bulge effective radius and central surface brightness vary alongside all primary disk parameters, while bar and ring positions and widths remain fixed.

\subsubsection{Parameter estimation}
We refine the automated initial decomposition via Maximum Likelihood Estimation (MLE). Prior to MLE optimization, the automatically identified model components for each object are visually inspected against both the surface brightness profile and the galaxy image, and adjusted manually if necessary. A second outer disk component is retained only if it satisfies a Bayesian Information Criterion test \citep[BIC;][]{1978AnSta...6..461S, 2007MNRAS.377L..74L},
\begin{equation}
    \Delta\mathrm{BIC} = \mathrm{BIC}_{1\,\mathrm{disk}} - \mathrm{BIC}_{2\,\mathrm{disks}} > 10,
\end{equation}
in addition to visual verification from the galaxy image. All model parameters are left free during the MLE optimization step to ensure a robust solution.

We adopt a log-likelihood function of the form:
\begin{equation}
    \ln\mathcal{L} = -\frac{1}{2} \sum_{i} \left[ \frac{\left(I_{\mathrm{model},i} - I_{\mathrm{data},i}\right)^2}{s_i^2} + \ln\left(2\pi s_i^2\right) \right],
\end{equation}
where
\begin{equation}
    s_i^2 = \sigma_{\mathrm{data},i}^2 + I_{\mathrm{model},i}^2 \cdot e^{2\ln f},
\end{equation}
and $\ln f$ is a free parameter introduced to account for potential underestimation of photometric uncertainties \citep[e.g.,][]{emcee_paper}.\footnote{\url{https://emcee.readthedocs.io/en/stable/}} Minimization is performed using the Sequential Least Squares Programming (\textsc{SLSQP}) algorithm.

Finally, parameter uncertainties for the S\'{e}rsic bulge and exponential disk components are estimated via Markov Chain Monte Carlo (MCMC) sampling using \textsc{emcee} \citep{emcee_paper}, as implemented in \textsc{LMFIT} \citep{newville_2025_16175987}, following the methodology of \citet{2025A&A...702A..42G}. The MCMC chains are run with a burn-in phase of 250 steps, followed by 2,500 production steps thinned by a factor of 25 (with \texttt{nan\_policy=`omit'} and \texttt{float\_behavior=`posterior'}). Parameter uncertainties are defined as the 16th and 84th percentiles of the marginalized posterior distributions. The exact mathematical expressions for each profile component are provided in Appendix~\ref{app:formulas}.

\subsubsection{Edge-on galaxies}

For 10 galaxies observed close to the edge-on orientation (see Appendix \ref{appendix:table}), we estimated the radial scale length of the low-surface-brightness disk using the photometric analysis presented in \cite{Saburova2024ApJ...973..167S}.
The radial and vertical light distributions were analyzed separately by extracting cuts parallel and perpendicular to the disk mid-plane and fitting them with projected edge-on exponential disk models. 
Here, however, our goal is limited to deriving a characteristic radial scale length, and therefore, we did not perform a separate analysis of vertical slices along the disk. 
We fitted the surface-brightness distribution with the standard projected edge-on exponential disk model integrated along the line of sight,
\begin{equation}
    I_R(R,z) \propto \frac{R}{h} K_1\!\left(\frac{R}{h}\right),
\end{equation}
where $K_1$ is the modified Bessel function of the second kind and first order, and $h$ is the radial scale length of the disk.

The fit was performed directly on the galaxy images after masking the central bulge-dominated region and contaminating fore- and background sources.
The full details of the original methodology and its application to edge-on galaxies are given in \cite{Katkov2019MNRAS.483.2413K, Saburova2024ApJ...973..167S}.

\section{Results}\label{results}

\subsection{Photometry analysis}
We performed photometric analysis of 148 gLSB candidate galaxies, out of which 10 were analyzed separately from the others due to their edge-on orientation.
We estimated the uncertainties for the S\'ersic bulge and exponential disk parameters of moderately inclined majority of the galaxies via Markov Chain Monte Carlo (MCMC) sampling, with reported values representing the 16th and 84th percentiles of the marginalized posterior distributions. If the parameter distribution for a bulge failed to show a clear maximum, those bulge parameters were excluded from subsequent analysis.

Table \ref{tab:master_columns} gives the parameters of all 148 gLSBG candidates analyzed in this paper. For the column descrition please refer to Appendix \ref{appendix:table} .

The resulting parameters of the disks are plotted in Fig.~\ref{fig:data_sample}, where we compared the central surface brightness and exponential scale length. We show all the gLSBGs candidates found in the 120 sq. deg. HSC footprint and other parts of the sky. We compare the parameters with those found in the literature and according to the results of the simulations (EAGLE, TNG100). The diagonal dotted line shows the diffusion criterion used to select gLSBGs in other papers \citep{Sprayberry1995}. As one can see from the figure, the diffusion criterion can also select the dwarf LSB galaxies \citep[see e.g. NGC1140][]{2025A&A...700A..56B}. In this work we focus specifically on giant systems, therefore the direct size and central surface brightness criteria adopted in this work are more efficient for selecting gLSBGs than the diffusion criterion. They are shown by red dashed lines. As one can see from Fig.~\ref{fig:data_sample}, not all TNG100 models selected using \HI data satisfy our criteria, since their disks are smaller than the adopted thresholds.

\begin{figure}
    \centering
    \includegraphics[width=\hsize]{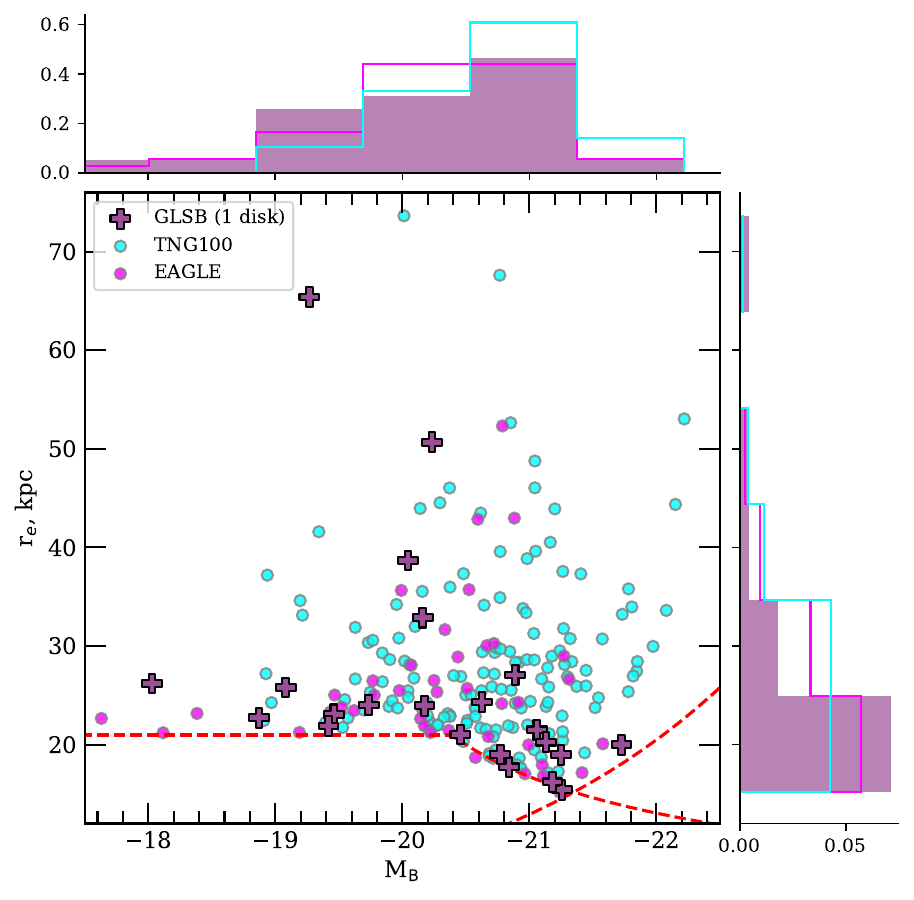}
    \caption{Effective radius versus absolute B-band magnitude for the gLSB disks inside the studied volume of this work (crosses), compared with model galaxies from the EAGLE and TNG100 simulations (circles). Dashed lines indicate the selection criteria adopted in this study. The top and side sub-panels show the distributions of observed and simulated galaxies along each parameter, with observed galaxies represented by the filled histograms.}
    \label{fig:sim}
\end{figure}

To make further progress in our understanding of the nature of gLSBGs disks, we also compared their effective radii with the absolute B-band magnitude together with EAGLE and TNG models in Fig.~\ref{fig:sim}. The conversion to the B-band was performed using the following transformation from \citep{Jester2005}: $m_B = m_g + 0.39(g-r) + 0.21$. For this comparison, we only used gLSBGs within the studied volume fitted with a single disk component to retain consistency with the \cite{Zhu23} as well as EAGLE \citep{vol_dens2023} analysis. The side panels demonstrate the PDF distributions by both parameters. The histograms with cyan and magenta contours correspond to TNG100 and EAGLE, respectively. One can see that TNG100 models have systematically higher disk luminosities in comparison to EAGLE models and observed gLSBGs. 

The measured bulge S\'ersic indices for the gLSBGs lie predominantly in the range $n \sim 1 - 2.5$.

\begin{figure*}
    \centering
    \includegraphics[width=\hsize,trim={0 0 0 0},clip]{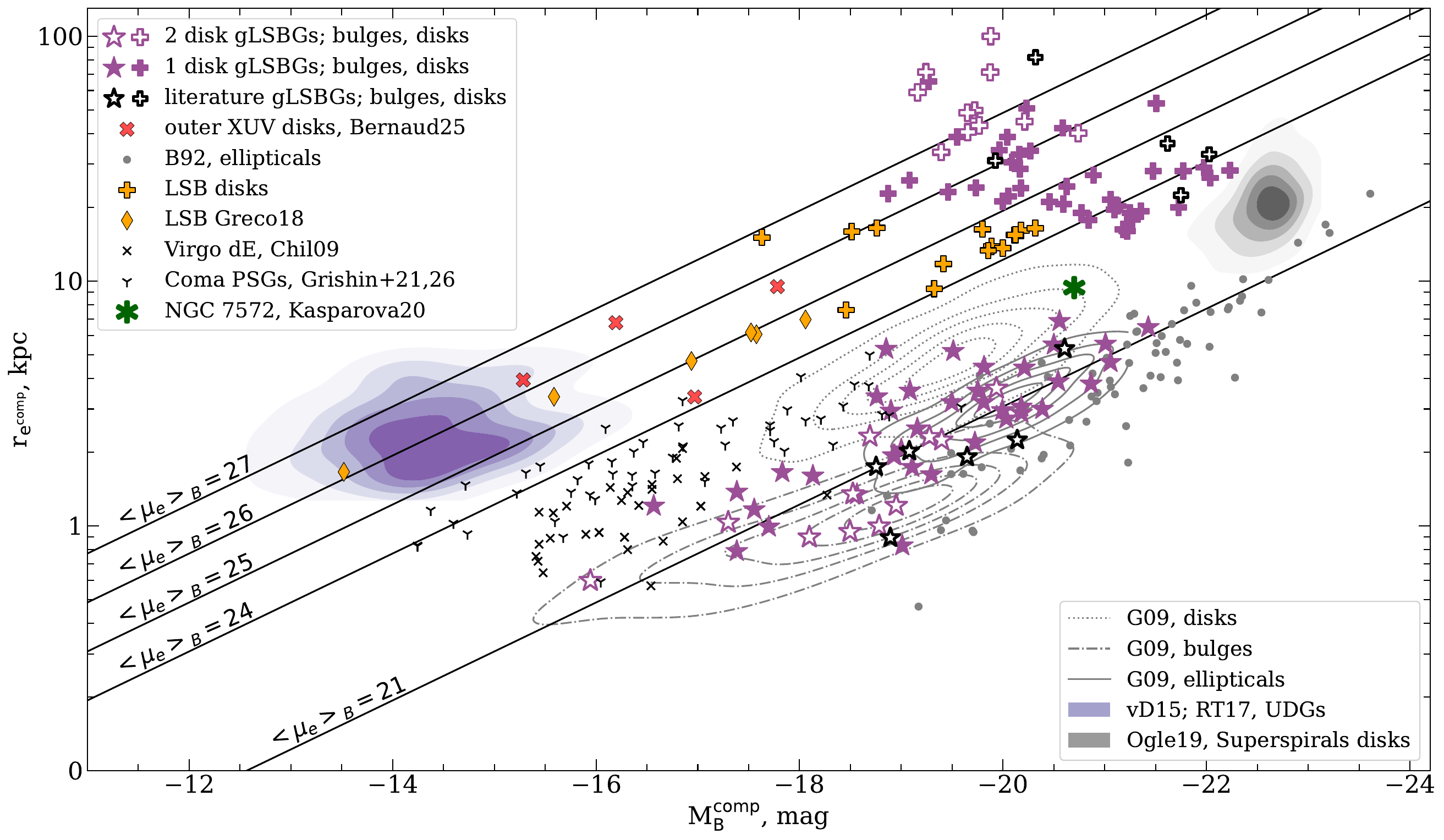}
    \caption{Effective radius versus absolute B-band magnitude for the bulge and disk components of gLSBGs, including both the sample presented in this work and literature data. For comparison, various classes of galaxies and galaxy components from the literature are also shown, including ultra-diffuse galaxies (UDGs), superspirals, dwarf ellipticals (dEs), elliptical galaxies, classical and pseudo-bulges, XUV disks, high-surface-brightness (HSB) disks, and moderately sized LSB galaxies. The lines correspond to the constant mean surface brightness in the B band in mag arcsec$^{-2}$ \citep{vanDokkum2015}.}
    \label{fig:Mre}
\end{figure*}
\nocite{2017MNRAS.468.4039R, 2009Sci...326.1379C, 2021NatAs...5.1308G}

In Fig. \ref{fig:Mre} we compare the position of gLSBGs bulges (stars) and disks (pluses) with the position of other types of galaxies on the absolute B-band magnitude versus effective radii diagram. In this diagram, we also show the lines of constant mean surface brightness and the locus of bulges and elliptical galaxies according to \citet{Gadotti2009, Bender1992}. LSB galaxies of moderate sizes are shown by yellow symbols (pluses -- ones identified in this study and diamonds from \citep{Greco2018}). We also show the position of extended ultraviolet (XUV) disks \citep{2025A&A...700A..56B} and super spirals \citep{Ogle2019}. As one can see, gLSBGs disks lie on the continuation of the trend formed by ultra diffuse galaxies (UDGs), moderate-size LSB galaxies and XUV disks, while dwarf elliptical (dE) and post-starburst (PSG) galaxies lie below the trend. One can see that the disks of superspiral galaxies lie on the continuation of the trend formed by late-type galaxy disks and have systematically lower effective radii and higher luminosities than the gLSB disks. The bulges of gLSBGs behave differently on the diagram -- some of them are occupying the regions of classical and pseudo-bulges. A significant part of the gLSBGs bulges lies in the area of elliptical galaxies.

We inspected deep images of our gLSBGs and revealed that distinct part of them are composed of a central HSB component surrounded by an extended, low-surface brightness disk. These central HSB components occupy the bright end of distributions defined by classical elliptical galaxies. The galaxies with brighter disks on average did not show such prominent bulges. This feature lead us to exploring the connection between the inner component and the disk of gLSBGs. In Fig. \ref{fig:Minner} we present diagrams correlating exponential disk structural parameters $\mu_0$ and $h$ with the total absolute magnitude of galaxy inner part. For literature data on high surface brightness (HSB) galaxies we used data from \citep{1996A&A...313...45D} and \citep{2003ApJ...582..689M}. For galaxies with a single disky component fit (filled purple triangles) the inner part is the bulge (Sersic component), for the ones with two disks (empty purple triangles) it is the inner late-type galaxy (Sersic+smaller disk). We observe trends for gLSBGs: ones with more luminous central components exhibit fainter, bigger and flatter giant LSB disks.

To compare the gLSBGs with XUV galaxies we performed a visual inspection of the Galaxy Evolution Explorer~\citep[GALEX;][]{2005ApJ...619L...1M} images of the confirmed gLSBGs. Three members of our team performed the inspection, assigning each galaxy with GALEX detection to one of three categories: with central, extended, or no UV features ('c', 'e' or 'n' in Table \ref{tab:master_columns} As follows from the table most of gLSBGs have extended UV features.

\subsection{Spectroscopic and kinematic properties}
We present the gLSBG central stellar velocity dispersions in Table \ref{tab:master_columns}. Due to their faint nature, LSB disks were not detected in the long-slit follow-up observations, and archival fiber-based spectroscopic surveys primarily captured the central regions of these systems. We also compare the central stellar velocity dispersions with the absolute B-band magnitude of the selection of gLSBG bulges -- the so-called Faber-Jackson relation~\citep{1976ApJ...204..668F} in Fig. \ref{fig:FJ}, where we also demonstrate the regions occupied by elliptical galaxies and bulges. As shown in Fig. \ref{fig:FJ}, many gLSBGs in the Faber-Jackson scaling relation  occupy the locus of elliptical galaxies.

\subsection{gLSBG number count and volume density estimate}
\label{sec:voldens_results}

From our candidate sample of 148 galaxies, we determine gLSBG number counts under two central surface brightness criteria: (i) $\mu_{0, B}^{\text{disk}} \ge 22~\text{mag~arcsec}^{-2}$, following standard LSB galaxy literature, and (ii) $\mu_{0, B}^{\text{disk}} \ge 23~\text{mag~arcsec}^{-2}$, following \citet{vol_dens2023}. Both criteria apply the same physical size threshold ($4h_{\mathrm{d}} \ge 50~\mathrm{kpc}$ or an isophotal radius $R_{28,B} \ge 50~\mathrm{kpc}$), yielding $60$ and $40$ gLSBGs, respectively.

A subsample of these candidates forms a volume-complete set within a comoving volume of $9.8 \times 10^5\,\mathrm{Mpc}^3$. Based on this volume-complete sample, we derive updated gLSBG volume densities of $(3.6 \pm 0.6) \times 10^{-5}\,\mathrm{Mpc}^{-3}$ for the standard literature threshold ($\mu_{0, B}^{\text{disk}} \ge 22~\text{mag~arcsec}^{-2}$) and $(2.7 \pm 0.5) \times 10^{-5}\,\mathrm{Mpc}^{-3}$ for the stricter \citet{vol_dens2023} threshold ($\mu_{0, B}^{\text{disk}} \ge 23~\text{mag~arcsec}^{-2}$). For comparison, applying the diffuseness index  selection criterion of \citet{Sprayberry1995} yields a volume density of $(5.4 \pm 0.7) \times 10^{-5}\,\mathrm{Mpc}^{-3}$.

To quantify the relative abundance of gLSBGs within their luminosity range ($1.4 \times 10^{10} < L_g / L_\odot < 1.37 \times 10^{11}$), we compare these space densities against the $g$-band field galaxy luminosity function parameters from \citet{2003ApJ...592..819B}. Under the standard $\mu_{0, B}^{\text{disk}} \ge 22~\text{mag~arcsec}^{-2}$ criterion, gLSBGs account for approximately 1 in every 4,000 galaxies in this luminosity regime.

To investigate discrepancies with theoretical models, we applied uniform selection criteria ($\mu_{0, B}^{\text{disk}} \ge 22~\text{mag~arcsec}^{-2}$ alongside $4h_{\mathrm{d}} \ge 50~\mathrm{kpc}$ or $R_{28,B} \ge 50~\mathrm{kpc}$) to cosmological simulations. This yields $150$ gLSBGs in TNG100 \citep[using the sample selection of][]{Zhu23} and $43$ in EAGLE (using the sample from \citet{vol_dens2023}), corresponding to volume densities of $(14.6 \pm 1.2) \times 10^{-5}\,\mathrm{Mpc}^{-3}$ and $(4.1 \pm 0.7) \times 10^{-5}\,\mathrm{Mpc}^{-3}$, respectively.

\section{Discussion}\label{discussion}
We identified a population of 61 gLSBGs using a robust photometric decomposition method homogenoeusly applied to a sample of 148 candidate galaxies. This yielded a complete sample of gLSBGs within a volume out to $z=0.1$ and additionally enabled the population-based approach to the analysis of gLSBGs with implications on their possible formation scenarios as a class of galaxies as well as comparison with other galaxy types.

\subsection{Implications for the gLSBG formation scenarios} 
\begin{figure}
    \centering
    \includegraphics[width=\hsize]{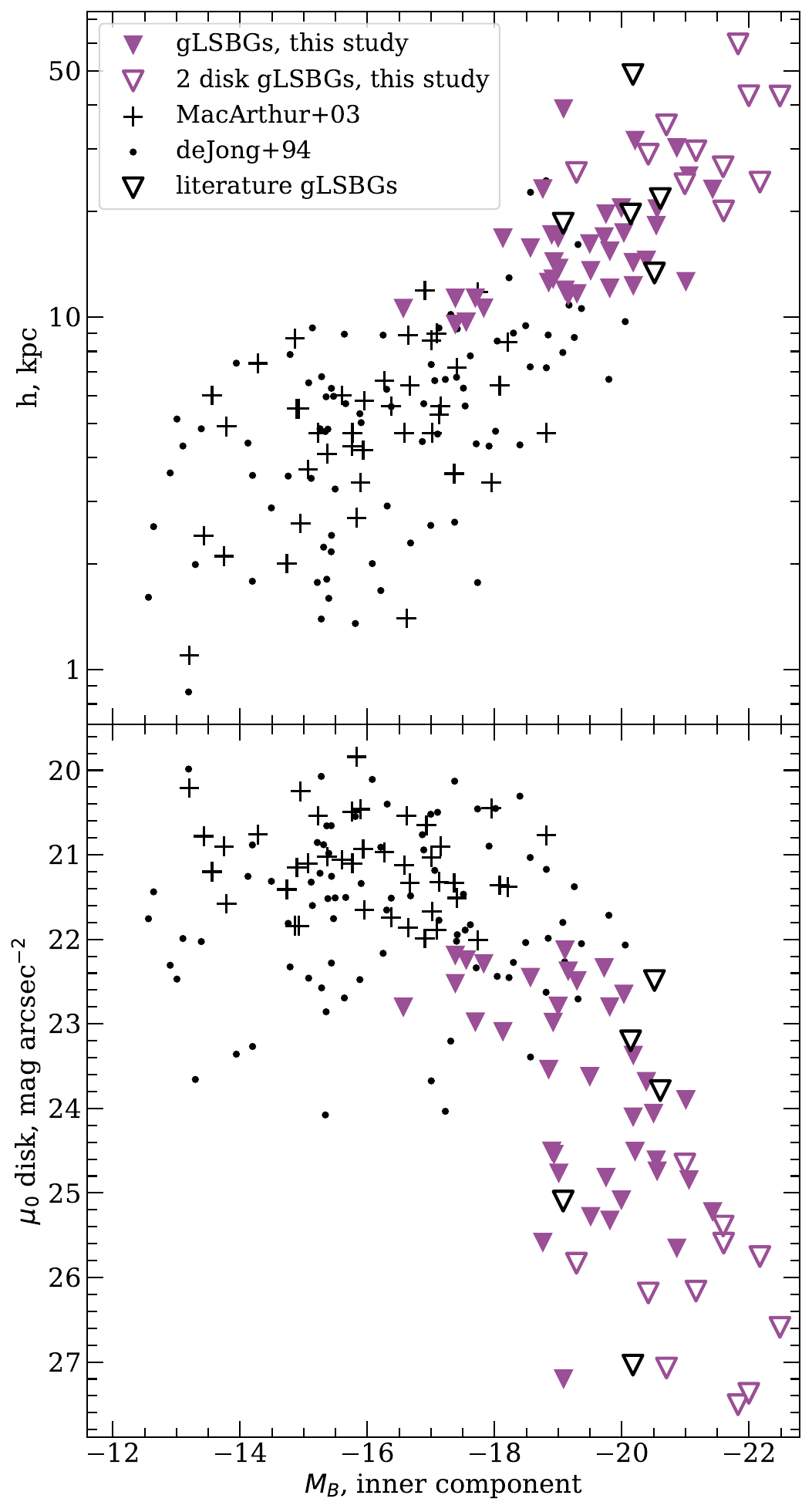}
    \caption{Absolute B-band magnitude of the galaxy inner component vs. exponential scalelength of the disk (upper panel) and disk central surface brightness (lower panel). Filled purple triangles show gLSBGs with single component disk fit. In this case inner component is the Sersic profile. Empty purple triangles correspond to the gLSBGs with double component disk fit. In this case inner component magnitude corresponds to the Sersic+smaller disk profile. We also give literature data on late-type galaxies for comparison.}
    \label{fig:Minner}
\end{figure}

\begin{figure}
    \centering
    \includegraphics[width=\hsize]{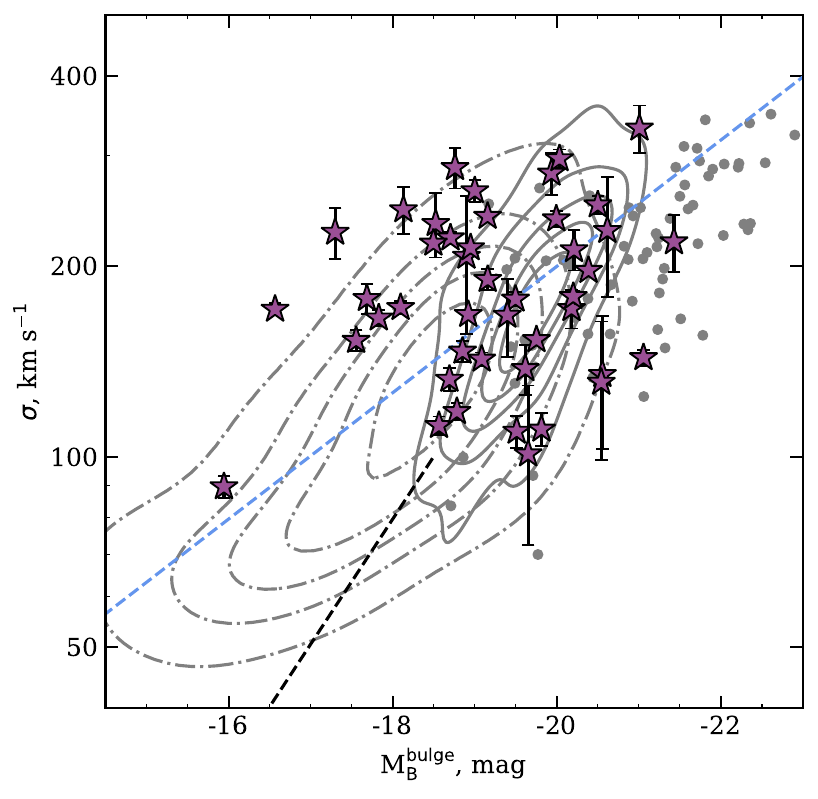}
    \caption{Central stellar velocity dispersion versus B-band absolute magnitude of a bulge. The position of gLSBGs is given by stars. For comparison, we also show elliptical galaxies, classical and pseudo- bulges. Velocity dispersions for \citep{Gadotti2009} are taken from RCSED \citep[\url{rcsed.sai.msu.ru},][]{Chilingarian2017}.}
    \label{fig:FJ}
\end{figure}

Based on population-wide statistics we propose gLSBG formation scenarios to drastically differ from hierarchical evolution and inside-out growth that dominate the assembly of late-type galaxies. These exhibit exponential stellar surface brightness profiles, which were built during gigayears of the inside-out formation \citep{1976MNRAS.176...31L, 1991ApJ...379...52W, 1998MNRAS.295..319M}. Conversely, disk structural parameters of gLSBGs presented in this work (i) scale with the luminosity of the inner galaxy component and (ii) present flatter and fainter disk profiles as the galaxy gets bigger (see Fig. \ref{fig:Minner}). The flat disk surface brightness profile suggests a rapid formation process with star formation burst at all radii simultaneously \citep{2005MNRAS.363.1299O}. 

The HSB part can either be (i) a pre-formed bulge/elliptical galaxy (see filled star symbols in Fig.~\ref{fig:Mre}), or (ii) a pre-exisitng late type galaxy. The latter ones show to have the largest and flattest gLSB disks (see empty triangles in Fig.~\ref{fig:Minner}) with Mailn~1 being a prime example. \citet{2016PASJ...68....2S} argue that moremassive central HSB parts of disky galaxies are associated with more massive dark matter haloes. Those create a deeper graviational potetial well that is able to host an extended disk. This can explain the trend in Fig.~\ref{fig:Minner}, where more luminous (and hence more massive) central HSB parts are surrounded by the largest gLSB disks.

Several in-depth gLSBG studies explored the origin of the gas that led to formation of gLSB disks. They reported gLSB disks which exhibit flat radial gas-phase metallicity gradients \citep{Saburovaetal2018,Junais2024, saburovaetal2026}. This contradicts the inside-out formation scenario which should produce negative gas-phase metallicity gradients \citep{2017MNRAS.467.1154S, 2021MNRAS.502.5935S}. It also suggests the gLSB disk formation by wet mergers and/or accretion of cooled down gas from the hot corona \citep{2013MNRAS.434.1531F}.

\subsection{The updated gLSBG volume density estimate}

We update the volume density estimate of gLSBGs by revising the total number of systems in the  volume from \citep{vol_dens2023} and reassess the associated uncertainties. Our updated sample includes a total of 28 gLSBGs used for the volume density calculation using the same photometric criteria of $\mu_{0, B}^{\text{disk}} \ge 23 $ mag arcsec$^{-2}$; and $4h_{\text{d}}\ge50$ kpc or $R_{28} \ge 50$ kpc. This number is lower than 37 galaxies that reported in \citep{vol_dens2023}. The discrepancy arises primarily from masking flattened out tails in radial surface brightness profiles, which previously were artificially increasing the exponential disk scalelength.

Our revised volume density of $(2.9 \pm 0.5) \times 10^{-5}\,\mathrm{Mpc}^{-3}$ (see Section~\ref{sec:voldens_results}) remains significantly lower than theoretical predictions. The initial gLSBG selection in \citet{Zhu23} was based on $\mathrm{H\,\textsc{i}}$ disk radii; however \HI data were unavailable for the majority of our galaxies making it impossible to apply such criteria. According to Figures ~5 and~8 from \citet{Zhu23} and Fig. \ref{fig:data_sample} from this work, not all the gLSBGs from \citet{Zhu23} satisfy our optical photometric criteria. This narrows the gap between the gLSBG volume density estimates, but still leaves a 3.8-fold discrepancy between the volume density estimates.

Neither EAGLE nor TNG100 synthetic images take into account dust extinction. The dust should lower surface brightnesses, increase disk scale lengths, and decrease disk luminosities. Nevertheless, using mock images generated with the Monte Carlo radiative transfer code \textsc{SKIRT} \citep{2015A&C.....9...20C} \citet{Kulier2020} showed that the impact of dust for LSB systems is relatively small. 
According to this study, for the brightest gLSBG models in the EAGLE sample, the dust extinction correction in the mean surface brightness within the disk radius is expected to be only about 0.15 mag\,arcsec$^{-2}$ or below (See Fig. A2 in \citet{Kulier2020}). 
Also \citet{Kulier2020} demonstrated that the radii of the 28th B-band isophote in LSB galaxies measured with and without dust are usually very similar. Therefore, taking into account the dust in the simulations would not significantly increase the number of galaxy models satisfying our photometric selection criterion.

To further investigate the differences in gLSBG volume densities in TNG100 and observations, we compare the observed gLSBG population with samples drawn from the TNG100 and EAGLE simulations.  The resulting stellar mass-size distributions are shown in Fig.~\ref{fig:sim}. Visually, TNG100 exhibits a pronounced excess of gLSBGs with higher total brightness compared to both the observed volume-complete sample and EAGLE. This points to underlying  fundamental discrepancies between the observations and the TNG100 simulation. It may suggest that uncertainties, e.g., in the implemented feedback prescriptions in TNG100, promote the formation of excessively large disks that are unlikely to remain intact in reality for prolonged periods of time.

\subsection{gLSBGs, Superluminous Spirals and XUV galaxies}

gLSBGs and super-spiral galaxies host some of the largest disks among late-type galaxies. Their disks, however, differ drastically in mean surface brightness. Fig.~\ref{fig:Mre} shows that super-spirals follow the scaling trend established by normal disk galaxies in SDSS, whereas gLSBGs foccupy the larger end of LSB galaxies trend. Here, we evaluate whether the evolution of a typical gLSBG into a super-spiral is physically plausible.

To assess that, we consider a simple gas-to-stars conversion argument. A typical H\,\textsc{i} mass of a gLSB is $\sim 2\times10^{10}\,M_\odot$ \citep{2026ApJ...999..218L}. Adjusting by 1.4 with the standard correction for helium and metals \citep{2000ApJ...533L..99M}, this corresponds to a total gas mass of $\sim 2.8\times10^{10}\,M_\odot$. If all of this gas was converted into stars, assuming a mass-to-light ratio in the $g$ band of $M/L_g \simeq 0.5$--1 (for $g-r \approx 0.3$~mag), the $g$-band luminosity would increase only by $(2.8$--$5.6)\times10^{10}\,L_\odot$.

A typical absolute stellar magnitude of a super-spiral disk is $M_B \approx -22.5~\text{mag}$, whereas gLSBG disks span a wider range of luminosities with a characteristic value of $M_B \approx -20~\text{mag}$ (see Fig.~\ref{fig:Mre}). This magnitude offset corresponds to a luminosity difference of $\sim10^{11}\,L_\odot$. Thus, even in an extreme scenario where all available gas is converted into stars, typical gLSBGs remain significantly fainter than typical super-spirals. Furthermore, super-spirals retain significant gas reservoirs, at least in molecular form \citep{Lisenfeld2023}. Although their luminosity distributions overlap at the extreme ends, super-spirals are generally better supplied with star-forming material than gLSBGs. Converting the gas content of a gLSBG into stars could, however, yield a disk luminosity comparable to that of the super-S0 galaxy NGC~7572 \citep{Kasparova2020}. Nonetheless, as shown in Fig.~\ref{fig:Mre}, NGC~7572 has a significantly smaller disk size than a typical gLSBG.

A few gLSBGs at the bright end of the distribution, specifically those with $\mu_{0,B}^{\text{disk}}$ between $22$ and $23~\text{mag~arcsec}^{-2}$, overlap with the super-spiral population in Fig.~\ref{fig:Mre}. These systems could, in principle, evolve into super-spirals if their gas reservoirs are efficiently converted into stars.

We also compare gLSBGs to extended UV (XUV) disk galaxies. Malin~1 is a well-known XUV system, and \citet{2008ApJ...681..244B} similarly classify the gLSBGs Malin~2 and NGC~7589 as XUV disks, establishing an overlap between the two populations. Additionally, Fig.~\ref{fig:Mre} demonstrates that the outer disk components from double-exponential fits of XUV optical counterparts \citep{2025A&A...700A..56B} follow the classical LSB sequence.

Visual inspection of GALEX images reveals that a significant fraction of confirmed gLSBGs host extended UV emission, reinforcing the connection between XUV and LSB disks. Because GALEX survey depth is spatially inhomogeneous, additional XUV counterparts to LSB disks likely remain undetected. Furthermore, the volume density of XUV galaxies, $(1.5\text{--}4.2)\times10^{-3}\,\mathrm{Mpc}^{-3}$ \citep{2011ApJ...733...74L}, is orders of magnitude higher than that of gLSBGs, making XUV disks statistically likely companion structures to gLSBGs.

In summary, our observational findings demonstrate a clear physical connection between gLSB and XUV disks, whereas no such link exists for super-spirals. This suggests that while gLSBGs and XUV galaxies share common evolutionary pathways, super-spiral disks originate via distinct formation mechanisms. Confirming this scenario will require detailed environmental analyses alongside in-depth studies of individual systems.

\section{Summary}\label{conclusions}
In this work, we present a comprehensive study of a large sample of gLSBGs with the emphasis on structural properties of gLSBGs as a class of galaxies. Through visual inspection of deep photometric surveys, we identified a sample of 148 candidates based on their low-surface brightness morphological features. For 31 candidates lacking redshifts, we carried out follow-up spectroscopic observations. Of the 148 candidates, we confirmed $40$ galaxies to be gLSBGs based on photometric criterion for the disk with $\mu_{0, B}^{\text{disk}}>23$ mag arcsec$^{-2}$ in B band and $60$ with $\mu_{0, B}^{\text{disk}}>22$ mag arcsec$^{-2}$, with disk size larger than 50 kpc in both cases. This substantial sample of previously rather rare gLSBGs, together with homogeneous decomposition of their radial photometric profiles, allowed us to draw population-based conclusions for this type of galaxies.

We observe the gLSB disks are fainter and flatter for more luminous corresponding inner parts. This suggests a rapid star formation scenario in gLSB disks at all radii simultaneously to dominate the inside-out evolution as the disk size increases. Based on this and flat metallicity gradients from the previous in-depth studies of gLSBGs, we propose that the gLSB disks formed on short timescales with a likely contribution of wet mergers and gas accretion from the halo.

The accurate photometric decomposition of gLSBGs in the volume from \citet{vol_dens2023} together with those without previously known redshifts allowed us to update the volume density estimate of these galaxies:  $(3.6 \pm 0.6) \times 10^{-5}\,\mathrm{Mpc}^{-3}$ ($\mu_{0, B}^{\text{disk}} \ge 22~\text{mag~arcsec}^{-2}$) and $(2.7 \pm 0.5) \times 10^{-5}\,\mathrm{Mpc}^{-3}$ ($\mu_{0, B}^{\text{disk}} \ge 23~\text{mag~arcsec}^{-2}$). It is still 3.8 times smaller than the value obtained from the TNG100 gLSBG sample and is similar to the estimate from the EAGLE simulations. We argue that the discrepancy with TNG100 may be caused by the difference in the selected galaxy populations: the $M_B$ distributions shows that the observed and TNG100 gLSBG samples are different, while showing similarities with the EAGLE sample.

Finally, we compare the gLSBGs with super-spirals and XUV disks. We discuss a possible evolutionary connection between them and Superspirals, which has to be ruled out based on order of magnitude discrepancy in baryonic mass between a typical gLSBG and a typical super-spiral. However, our study confirms a certain similarity of the gLSBGs and XUV galaxies, as both types follow the LSB trend on the disk $M_{B}-r_{\text{eff}}$ diagram (see Fig. \ref{fig:Mre}) together with most of the gLSBGs from our sample hosting an extended UV component based on the visual inspection of the GALEX images.

\begin{acknowledgements}
IC's research is supported by the Telescope Data Center at the Smithsonian Astrophysical Observatory. IC also acknowledges the support from the NASA ADAP-22-0102 grant (award 80NSSC23K0493) and NASA XMM-Newton Data Analysis grant (award 80NSSC22K0389). We complement the use of LLM-based tools (ChatGPT, Claude, Gemini for finding appropriate wording and references in some cases, Grammarly for grammar and spelling convention checks) with critical thinking, as was proposed in \citep{2024arXiv240920252F}. This work made use of Astropy\footnote{http://www.astropy.org}, a community-developed core Python package and an ecosystem of tools and resources for astronomy \citep{astropy:2013, astropy:2018, astropy:2022}, particularly Photutils \citep{larry_bradley_2023_7946442}, and in addition Python package SciPy \citep{2020SciPy-NMeth}. 

\end{acknowledgements}

\bibliographystyle{aa}
\bibliography{bibliography}

\begin{appendix}
\section{Line spread functions of EFOSC2, FAST and TDS}
Here we present plots of the 2nd degree polynomial fits of the line spread functions of the EFOSC2, FAST and TDS setups. See Sections \ref{sect:efosc2}, \ref{sect:fast}, \ref{sect:tds} for the setup description. The pink triangles represent the measurements of individual arc/sky linewidths fitted with gaussian profiles.
\begin{figure}[ht]
    \centering
    \includegraphics[width=\hsize,trim={0 0 0 0},clip]{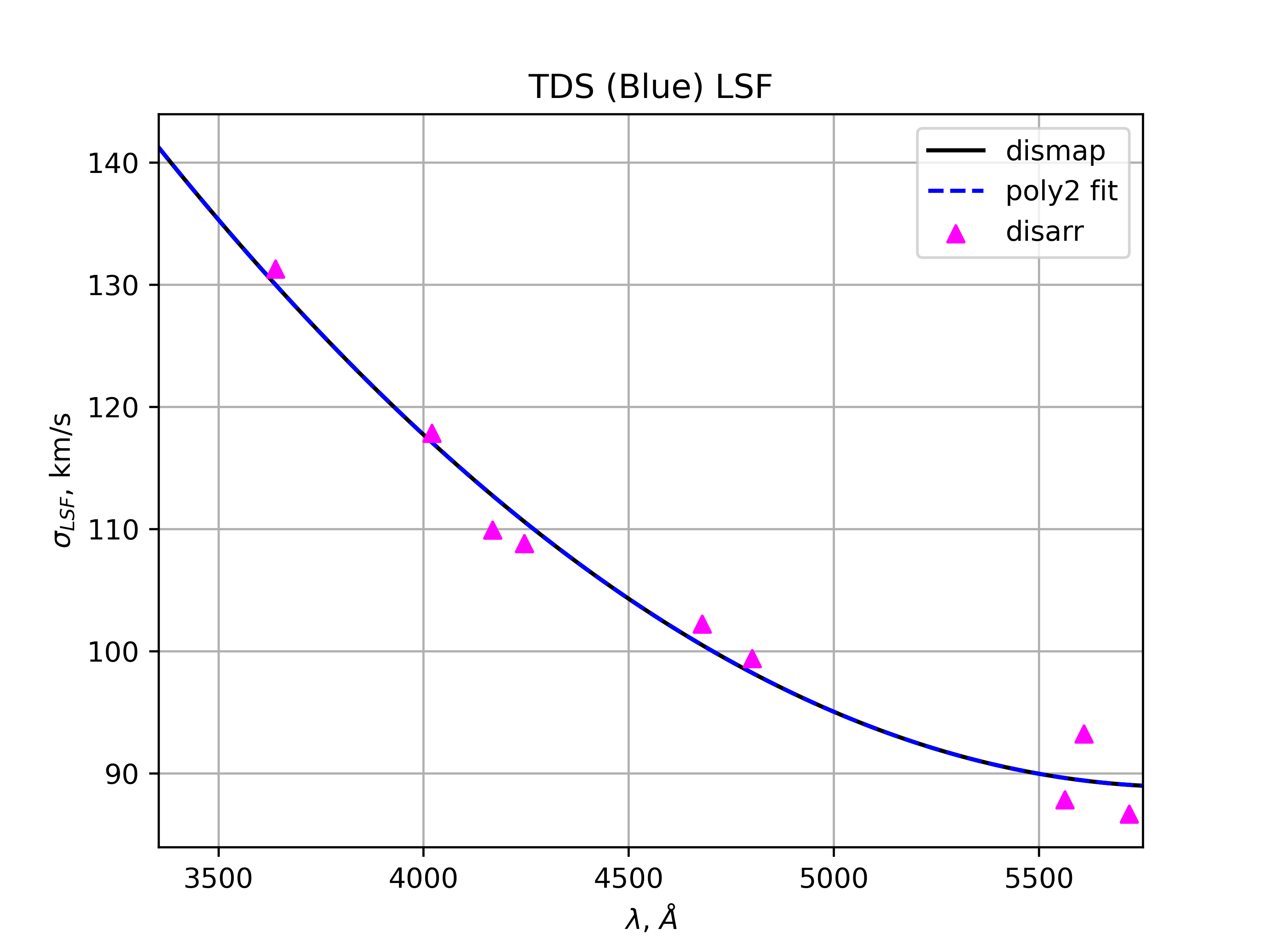}
    \includegraphics[width=\hsize,trim={0 0 0 0},clip]{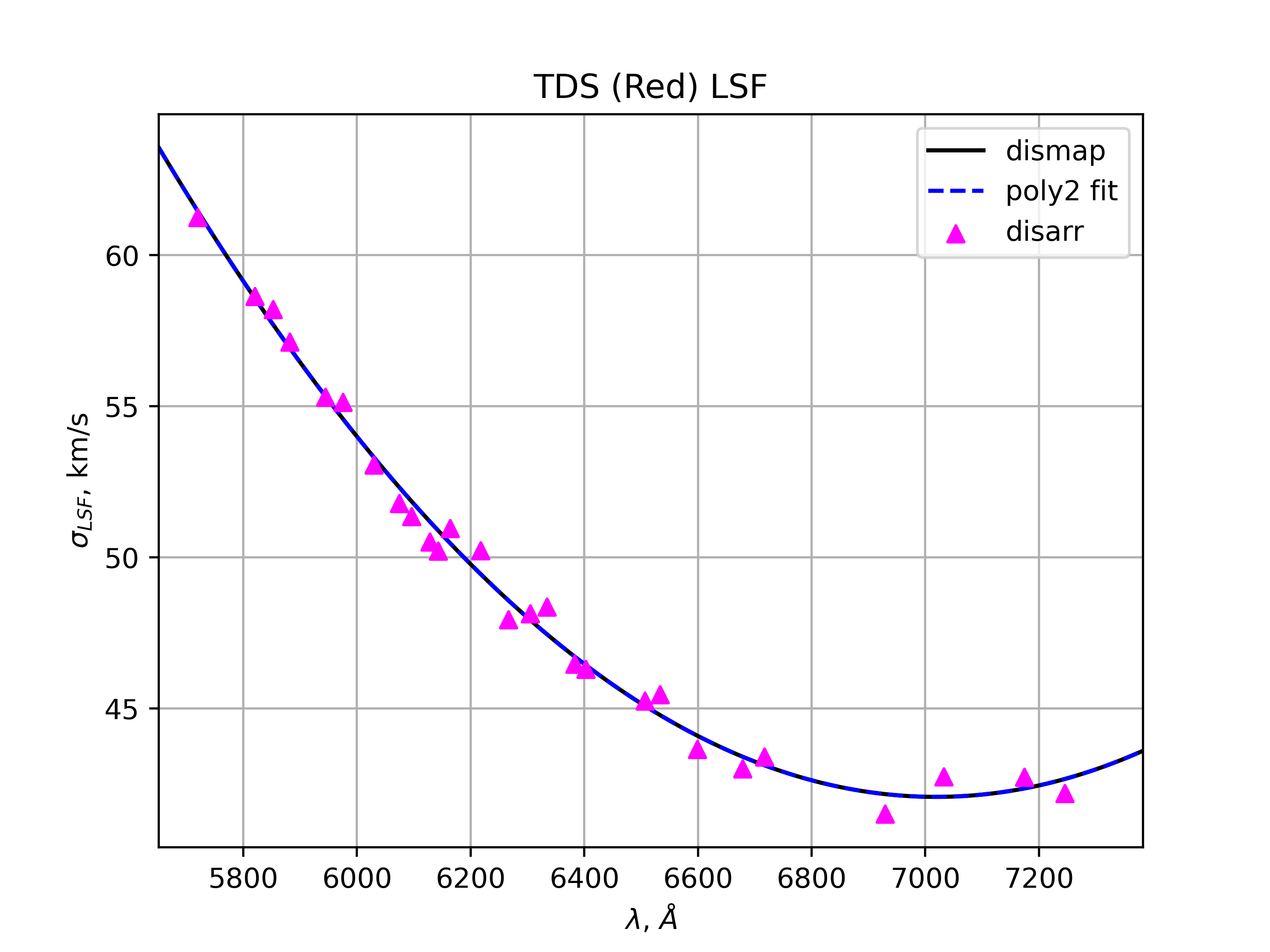}
    \caption{Line spread function of TDS setup, blue (B) and red (R) channels.}
    \label{fig:tds}
\end{figure}

\begin{figure}[ht]
    \centering
    \includegraphics[width=\hsize,trim={0 0 0 0},clip]{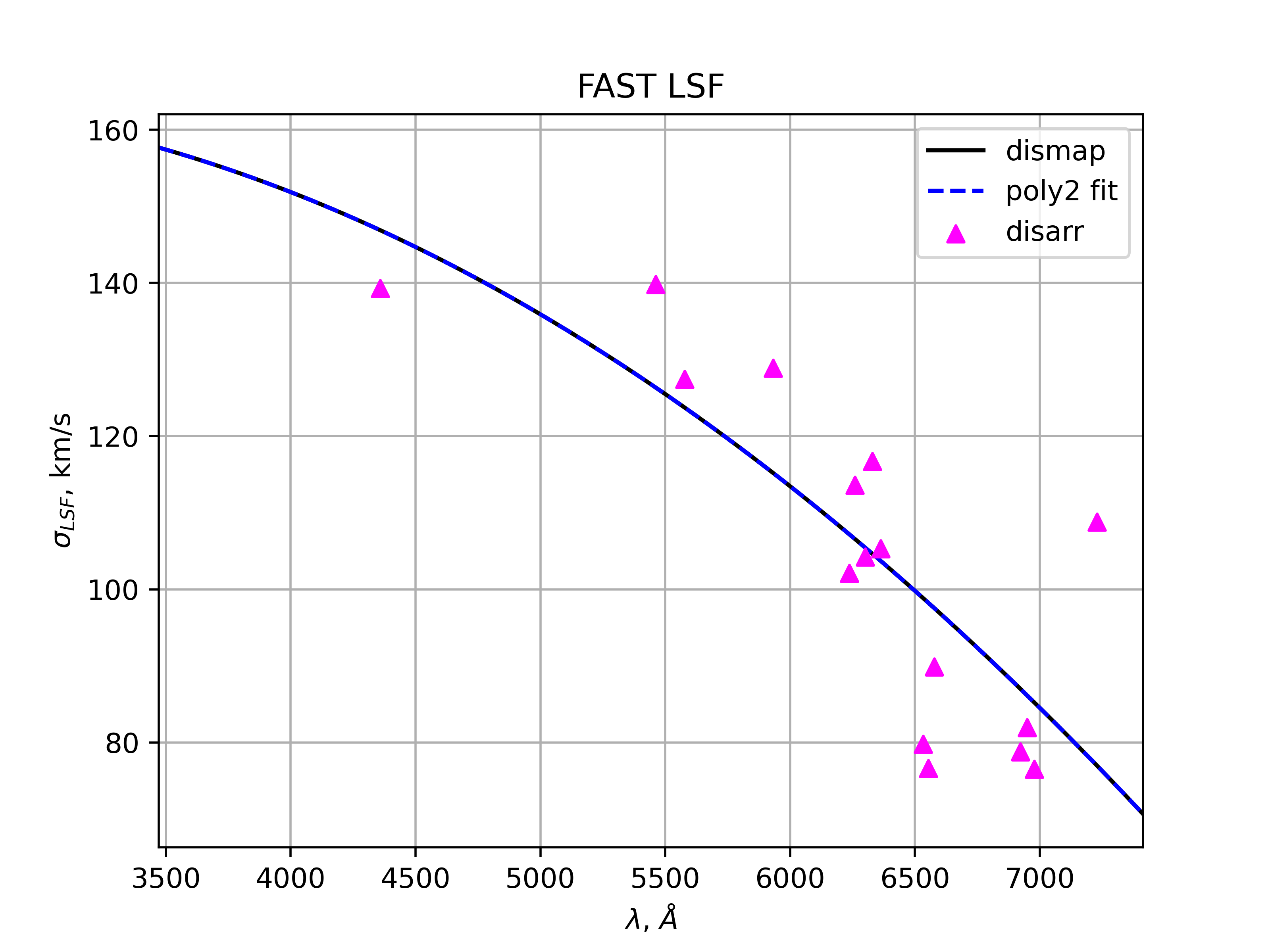}
    \caption{Line spread function of the EFOSC2 setup.}
    \label{fig:fast}
\end{figure}

\begin{figure}[ht]
    \centering
    \includegraphics[width=\hsize,trim={0 0 0 0},clip]{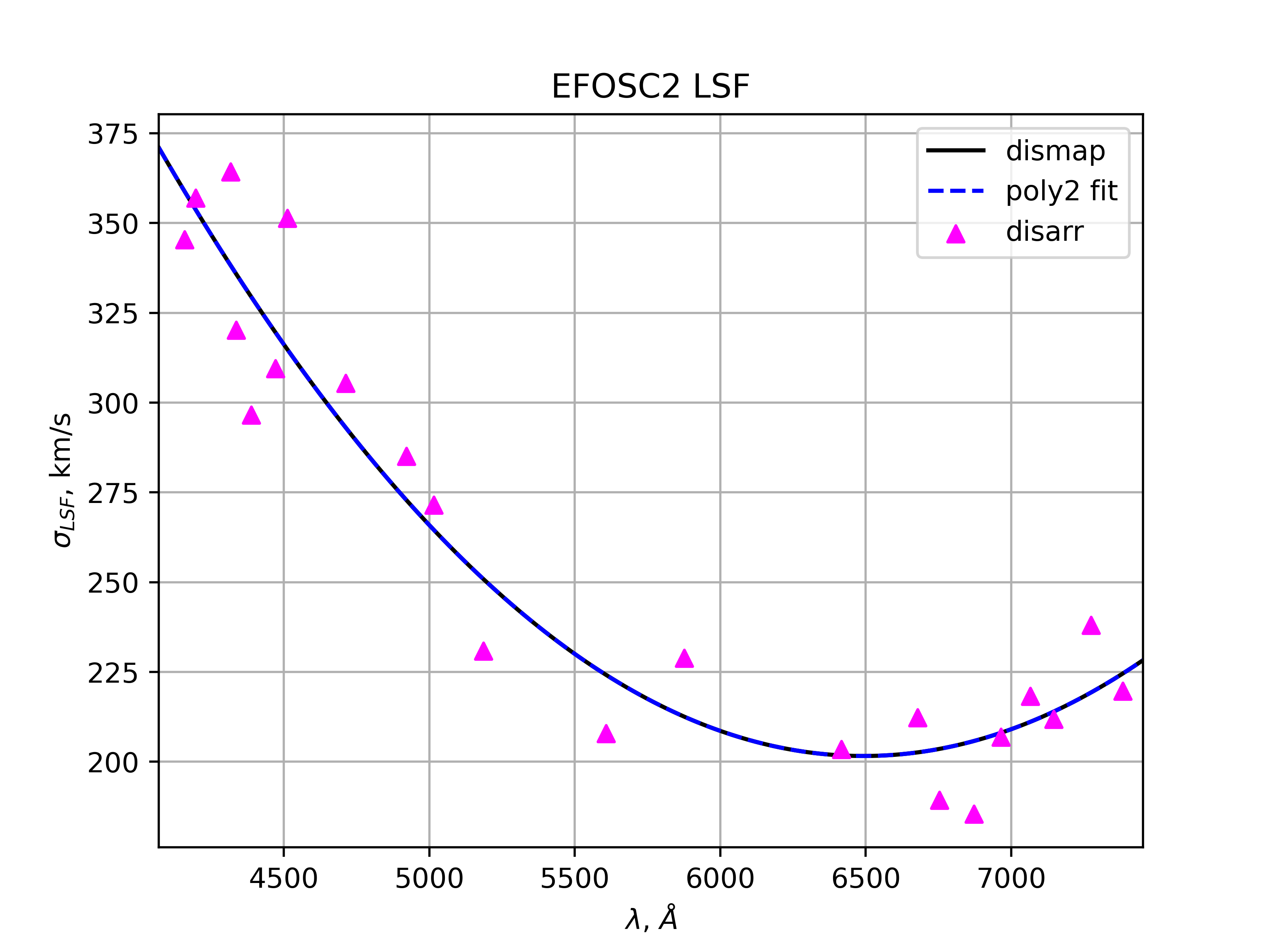}
    \caption{Line spread function of the FAST setup.}
    \label{fig:efosc}
\end{figure}

\section{Formulas of the component profiles}\label{app:formulas}
S\'ersic profile:
\begin{equation}
    I(r) = I_e \exp \left( -b(n) \left[ \left( \frac{r}{r_e} \right)^{\frac{1}{n}} - 1 \right] \right),
\end{equation}
Where $b(n) = 2n - \frac{1}{3} + \frac{4}{405n} + \frac{46}{25515n^2}$, and $r_e$, $I_e$ are effective radius and profile intensity at the effective radius respectively.

Exponential profile:
\begin{equation}
    I(r) = I_0 \exp \left( - \frac{r}{h} \right),
\end{equation}
where $I_0$ is the central brightness and $h$ is scalelength.

Fermi-Dirac profile:
\begin{equation}
    I(r) = I_0 \cdot \frac{1}{\exp\left(\frac{r - r_b}{h_{b}}\right) + 1},
\end{equation}
where  $I_0$ is amplitude of the distribution, $r_b$ is the truncation radius and $h_b$ is the truncation width.

Gaussian profile:
\begin{equation}
    I(r) = I_0 \cdot \frac{1}{\sqrt{2\pi} \cdot \text{w}} \exp\left(-\frac{(r - \mu_r)^2}{2 \text{w}^2} \right),
\end{equation}
where $I_0$ is central brightness of the profile, w is the width (standard deviation) and $\mu_r$ is the mean position.

\section{Data availability}\label{appendix:zenodo}
Follow DOI \texttt{10.5281/zenodo.22722142} to find (i) master parameter table (see \ref{appendix:table} for column descriptions); (ii) the data of the photometric profile decompositions for all 148 candidates; (iii)fits to the spectroscopic data of EFOSC2, FAST and TDS observations as well as DESI and 6dFGs archival spectra.

\onecolumn
\section{Parameter tables}\label{appendix:table}

\begin{longtable}{l l p{9cm}}
\caption{Columns of the master parameter table.} \label{tab:master_columns} \\
\hline
Column & Units & Description \\
\hline
\endfirsthead

\hline
Column & Units & Description \\
\hline
\endhead

\hline
\endfoot

\multicolumn{3}{l}{\textit{Identification and basic properties}} \\
name          & --            & Object identifier \\
z             & --            & Redshift \\
ra            & deg           & Right ascension (J2000) \\
dec           & deg           & Declination (J2000) \\
galex\_flag    & --            & GALEX evaluation flag for the object (\texttt{e} = extended, \texttt{c} = compact, \texttt{n} = not detected) \\
edge\_on\_flag  & --            & Boolean flag; \texttt{True} for galaxies in edge-on orientation, \texttt{False} for the main sample \\
\hline
\multicolumn{3}{l}{\textit{$g$-band structural decomposition}} \\
mu\_01\_g      & mag~arcsec$^{-2}$ & Central surface brightness of the primary exponential disk \\
mu\_01\_ep\_g   & mag~arcsec$^{-2}$ & Upper 1$\sigma$ uncertainty on mu\_01\_g \\
mu\_01\_em\_g   & mag~arcsec$^{-2}$ & Lower 1$\sigma$ uncertainty on mu\_01\_g \\
h1\_g          & kpc           & Scale length of the primary exponential disk \\
h1\_ep\_g       & kpc           & Upper 1$\sigma$ uncertainty on h1\_g (MCMC if edge\_on\_flag is \texttt{False})\\
h1\_em\_g       & kpc           & Lower 1$\sigma$ uncertainty on h1\_g (MCMC if edge\_on\_flag is \texttt{False})\\
mu\_02\_g      & mag~arcsec$^{-2}$ & Central surface brightness of the secondary exponential disk \\
mu\_02\_ep\_g   & mag~arcsec$^{-2}$ & Upper 1$\sigma$ uncertainty on mu\_02\_g (MCMC if edge\_on\_flag is \texttt{False})\\
mu\_02\_em\_g   & mag~arcsec$^{-2}$ & Lower 1$\sigma$ uncertainty on mu\_02\_g (MCMC if edge\_on\_flag is \texttt{False})\\
h2\_g          & kpc           & Scale length of the secondary exponential disk \\
h2\_ep\_g       & kpc           & Upper 1$\sigma$ MCMC uncertainty on h2\_g \\
h2\_em\_g       & kpc           & Lower 1$\sigma$ MCMC uncertainty on h2\_g \\
R27\_7\_g         & kpc           & Isophotal radius at $\mu_g = 27.7$~mag~arcsec$^{-2}$ \\
mue\_g         & mag~arcsec$^{-2}$ & Sérsic component surface brightness at the effective radius \\
mue\_ep\_g      & mag~arcsec$^{-2}$ & Upper 1$\sigma$ MCMC uncertainty on mue\_g \\
mue\_em\_g      & mag~arcsec$^{-2}$ & Lower 1$\sigma$ MCMC uncertainty on mue\_g \\
re\_g          & kpc           & Sérsic component effective radius \\
re\_ep\_g       & kpc           & Upper 1$\sigma$ MCMC uncertainty on re\_g \\
re\_em\_g       & kpc           & Lower 1$\sigma$ MCMC uncertainty on re\_g \\
n\_g           & --            & Sérsic index \\
n\_ep\_g        & --            & Upper 1$\sigma$ MCMC uncertainty on n\_g \\
n\_em\_g        & --            & Lower 1$\sigma$ MCMC uncertainty on n\_g \\
\hline
\multicolumn{3}{l}{\textit{$r$-band structural decomposition}} \\
mu\_01\_r      & mag~arcsec$^{-2}$ & Central surface brightness of the primary exponential disk \\
mu\_01\_ep\_r   & mag~arcsec$^{-2}$ & Upper 1$\sigma$ MCMC uncertainty on mu\_01\_r \\
mu\_01\_em\_r   & mag~arcsec$^{-2}$ & Lower 1$\sigma$ MCMC uncertainty on mu\_01\_r \\
h1\_r          & kpc           & Scale length of the primary exponential disk \\
h1\_ep\_r       & kpc           & Upper 1$\sigma$ MCMC uncertainty on h1\_r \\
h1\_em\_r       & kpc           & Lower 1$\sigma$ MCMC uncertainty on h1\_r \\
mu\_02\_r      & mag~arcsec$^{-2}$ & Central surface brightness of the secondary exponential disk \\
mu\_02\_ep\_r   & mag~arcsec$^{-2}$ & Upper 1$\sigma$ MCMC uncertainty on mu\_02\_r \\
mu\_02\_em\_r   & mag~arcsec$^{-2}$ & Lower 1$\sigma$ MCMC uncertainty on mu\_02\_r \\
h2\_r          & kpc           & Scale length of the secondary exponential disk \\
h2\_ep\_r       & kpc           & Upper 1$\sigma$ MCMC uncertainty on h2\_r \\
h2\_em\_r       & kpc           & Lower 1$\sigma$ MCMC uncertainty on h2\_r \\
mue\_r         & mag~arcsec$^{-2}$ & Sérsic component surface brightness at the effective radius \\
mue\_ep\_r      & mag~arcsec$^{-2}$ & Upper 1$\sigma$ MCMC uncertainty on mue\_r \\
mue\_em\_r      & mag~arcsec$^{-2}$ & Lower 1$\sigma$ MCMC uncertainty on mue\_r \\
re\_r          & kpc           & Sérsic component effective radius \\
re\_ep\_r       & kpc           & Upper 1$\sigma$ MCMC uncertainty on re\_r \\
re\_em\_r       & kpc           & Lower 1$\sigma$ MCMC uncertainty on re\_r \\
n\_r           & --            & Sérsic index \\
n\_ep\_r        & --            & Upper 1$\sigma$ MCMC uncertainty on n\_r \\
n\_em\_r        & --            & Lower 1$\sigma$ MCMC uncertainty on n\_r \\
\hline
\multicolumn{3}{l}{\textit{Kinematics}} \\
sig\_efosc     & km~s$^{-1}$   & Stellar velocity dispersion measured from EFOSC2 spectroscopy \\
esig\_efosc    & km~s$^{-1}$   & Uncertainty on sig\_efosc \\
sig\_fast      & km~s$^{-1}$   & Stellar velocity dispersion measured from FAST spectroscopy \\
esig\_fast     & km~s$^{-1}$   & Uncertainty on sig\_fast \\
sig\_tds      & km~s$^{-1}$   & Stellar velocity dispersion measured from TDS spectroscopy \\
esig\_tds     & km~s$^{-1}$   & Uncertainty on sig\_fast \\
sig\_survey    & km~s$^{-1}$   & Stellar velocity dispersion adopted from SDSS/6dFGS/DESI\\
esig\_survey   & km~s$^{-1}$   & Uncertainty on sig\_survey \\
survey         & --            & Name of the survey providing sig\_survey, i.e.  SDSS/6dFGS/DESI\\
\end{longtable}
\end{appendix}
\end{document}